\documentclass{article}

\usepackage{PRIMEarxiv}

\usepackage[utf8]{inputenc} 
\usepackage[T1]{fontenc}    
\usepackage{hyperref}       
\usepackage{url}            
\usepackage{booktabs}       
\usepackage{amsfonts}       
\usepackage{nicefrac}       
\usepackage{microtype}      

\usepackage{booktabs}   
\usepackage{multirow}   
\usepackage{microtype}  

\usepackage{lipsum}
\usepackage{fancyhdr}       
\usepackage{graphicx}       
\graphicspath{{media/}}     

\title{Synthesizing the Expert: A Validated Multimodal Dataset for Trustworthy AI-Assisted Swimming Coaching}

\author{  
  Ahmad Al-Kabbany\\
  Multimedia Interaction and Communication Lab \\
  Wearables, Biosensing, and Biosignal Processing Research lab \\
  Arab Academy for Science and Technology \\
  \texttt{alkabbany@ieee.org, alkabbany@aast.edu}
  \And
  Esraa Kassem\\
  Department of Sports Training and Movement Sciences, Faculty of Sports Sciences \\
  Alexandria University\\
  \texttt{esraa.m.kassem@alexu.edu.eg}
}

\begin{document}
\maketitle

\begin{abstract}
This research is primarily concerned with the critical problem of synthesizing and building a structured Retrieval-Augmented Generation (RAG) system for advanced AI applications in the domain of swimming. As the integration of Artificial Intelligence in sports science matures, its applications in swimming have become increasingly diverse, spanning from real-time technical coaching and talent scouting to comprehensive performance profiling and the dynamic personalization of training periodization. Within this landscape, RAG-based systems represent a pivotal advancement in Large Language Model (LLM) enhanced swimming analysis, as they allow for the grounding of generative outputs in authoritative domain knowledge, thereby ensuring that AI-generated advice is both contextually relevant and technically accurate.

Despite this potential, building robust RAG systems using only real-world aquatic data presents significant challenges, including data sparsity for critical edge cases, strict privacy and ethical constraints regarding athlete biometrics, and the high cost of manual expert labeling. To address these barriers, we propose a novel generative framework for the automated synthesis of an expert-level swimming corpus. This study leverages a comprehensive, multimodal knowledge base gathered across four dimensions: physiological data (e.g., $VO_{2}max$, $HRV$), physiological literature, high-frequency kinematic sensor data (e.g., 10-IMU network acceleration and gyroscope vectors), and unstructured domain expertise derived from elite coaching manuals.

Our proposed framework utilizes a multi-agent LLM architecture to synthesize 
a high-fidelity dataset of 1,864 validated ``Question-Context-Answer'' triplets 
— drawn from 1,914 drafts evaluated against 12 physiological soundness rules 
— meticulously mapping raw sensor spikes and physiological markers to 
evidence-based coaching interventions. By providing a structured, synthetic ground truth, this work establishes a foundational benchmark for trustworthy AI in aquatics. The outcomes of this research not only promise to enhance the reliability of automated coaching but also open a plethora of future directions in "Meta-Agent" development and cross-sport athletic profiling, ultimately bridging the gap between raw data engineering and practical sports science application.

\end{abstract}


\section{Introduction}
In the contemporary era of high-performance aquatics, the integration of wearable technology and physiological monitoring has moved from experimental novelty to an essential pillar of elite athlete preparation. The proliferation of multi-axial inertial measurement units (IMUs), often deployed in distributed sensor networks across the swimmer's body, now enables the capture of micro-movements with unprecedented temporal resolution \cite{morais2022wearables}. When coupled with real-time biometric tracking—measuring metrics such as heart rate variability (HRV), blood oxygen saturation via fNIRS, and metabolic markers—data sources even from regional-level swimmers have become remarkably pervasive \cite{keating2025longitudinal}. This technological ubiquity has transformed the swimming pool into a high-dimensional data environment, where every stroke cycle, turn phase, and physiological response is digitized into continuous multivariate time-series streams. Consequently, the primary challenge in swimming science has shifted from the difficulty of data acquisition to the management of a relentless data explosion that defines modern Olympic-cycle training \cite{muanescu2025big}.

While the volume of digitized swimming data has increased exponentially, the transition from raw data collection to actionable "prescriptive intelligence" has hit a critical knowledge bottleneck. The sheer scale of high-dimensional information—where a single training session can yield thousands of data points across sixty or more concurrent sensor channels—has become huge beyond the human cerebral capacity to interpret in real-time \cite{hammes2022artificial}. For a coach to manually reconcile subtle gyroscopic deviations in a swimmer's "catch" phase with fluctuating heart rate variability (HRV) and fatigue markers requires a level of multi-variate analysis that exceeds traditional observational methods. Consequently, a "semantic gap" has emerged: we possess the descriptive tools to know what occurred in the pool, yet we lack the automated, reliable systems necessary to prescribe how to optimize subsequent sets \cite{xu2025reshaping,pisaniello2024game}. Without an intelligence layer capable of synthesizing these disparate data streams, the potential for personalized, data-driven coaching remains largely untapped, trapped within a deluge of uninterpreted signals \cite{fitrianto2025systematic}.

This prescriptive failure is further compounded by a structural paradox at the heart of aquatic data science: despite the volumetric abundance of raw sensor streams, the data that truly matters — labeled, expert-validated, and actionable — remains critically scarce. While high-frequency signals from 10-sensor IMU networks are pervasive, the availability of high-quality, structured datasets — specifically those containing synchronized, expert-vetted annotations — is nearly non-existent in the public research domain \cite{patil2025ai}. This creates a profound volume-versus-utility imbalance where researchers possess massive logs of numerical vectors but lack the "ground truth" labels required to train prescriptive models. Moreover, the immense temporal and financial costs associated with manual expert labeling for complex biomechanical movements often result in a "sparse data" problem for critical edge cases, where models struggle to generalize beyond simple stroke classification \cite{puce2025role}.

Compounding this labeling challenge is a parallel ethical constraint that fundamentally limits data shareability. The acquisition and dissemination of elite swimming data are governed by heightened protections under global privacy legislation, most notably the General Data Protection Regulation (GDPR), which classifies athlete biometrics as a special category of sensitive personal data \cite{kwon2025athlete}. This legal framework imposes strict boundaries on the open sharing of real-world performance profiles, creating a significant barrier to reproducible, collaborative research. Consequently, the development of intelligent swimming systems has reached a plateau where the scientific community lacks a publicly available, expert-annotated benchmark — not merely due to the difficulty of labeling, but due to the fundamental impossibility of sharing the underlying raw data at all.

\begin{figure}[t]
    \centering
    \includegraphics[width=\textwidth]{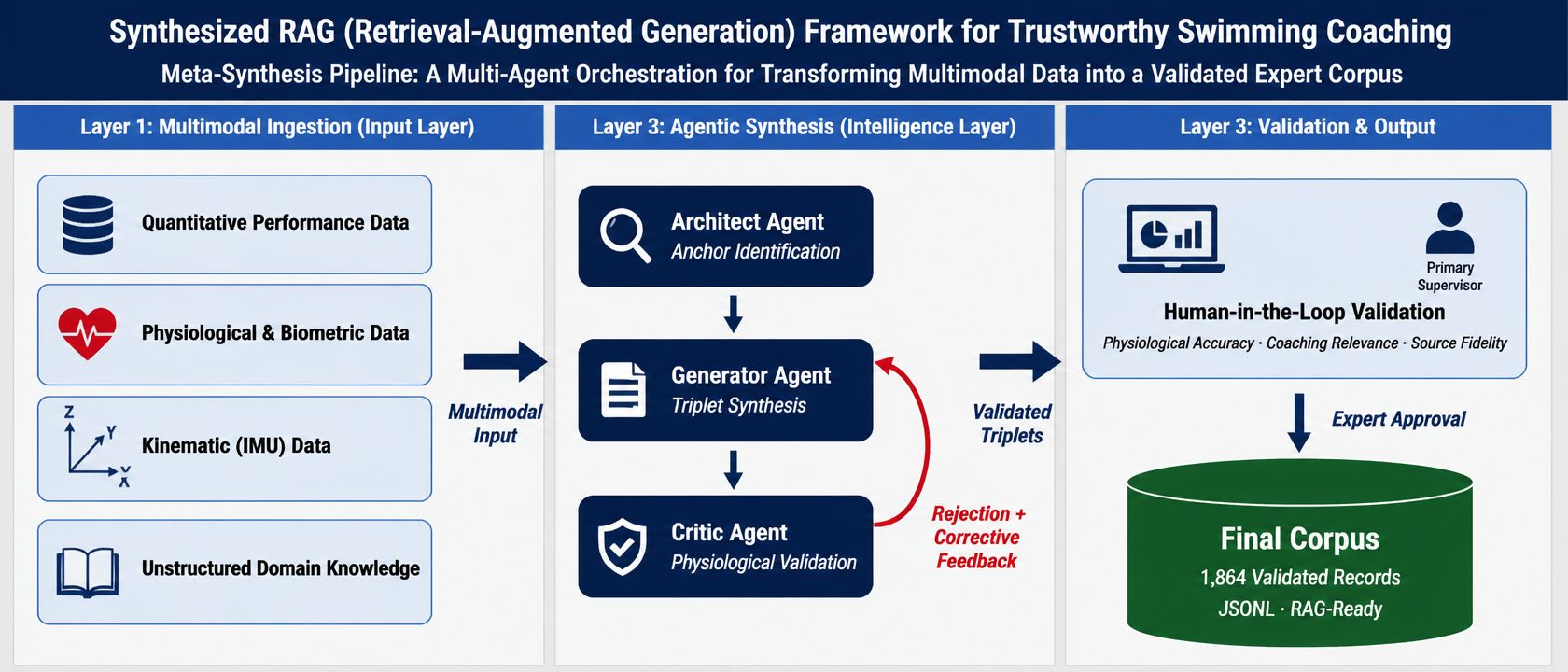}
    \caption{Proposed Meta-Synthesis RAG Pipeline for trustworthy swimming coaching. The framework illustrates the transformation of multimodal data (kinematic, physiological, and unstructured) through a multi-agent orchestration layer, resulting in a validated expert corpus of 1,864 records via human-in-the-loop validation. Please see text for more details.}
    \label{fig:framework_pipeline}
\end{figure}

To bridge this dual barrier of labeling scarcity and privacy constraint, we propose a novel generative framework for 'Synthesizing the Expert,' which leverages a multi-agent orchestration layer to create a high-fidelity, multimodal swimming corpus. Unlike static retrieval systems, our framework employs an agentic architecture — comprising specialized Architect, Generator, and Critic agents — that autonomously maps raw multivariate signals to authoritative domain knowledge \cite{wu2026personalized}. Grounded in a heterogeneous knowledge base spanning the Brunner et al. 10-IMU sensor network, physiological markers including $VO_{2}max$ and lactate thresholds, and elite coaching references such as the Stager Handbook and established periodization protocols, the 'Architect Agent' first identifies performance anchors by correlating biomechanical deviations with their corresponding physiological rules. These anchors are then processed by the 'Generator Agent' to synthesize 
1,864 validated 'Question-Context-Answer' triplets — drawn from 1,914 drafts 
evaluated against 12 physiological soundness rules — each grounded in a 
verifiable source rather than generative assumption. To ensure scientific rigor, the 'Critic Agent' enforces physiological soundness by cross-referencing all generated coaching prescriptions against established recovery and periodization protocols, explicitly rejecting outputs that conflict with documented fatigue or adaptation markers. By transforming sparse and privacy-constrained biometric data into a structured, anonymized knowledge base, this framework provides the first scalable and trustworthy foundation for RAG-based coaching applications in competitive aquatics \cite{kechun2026retrieval,bunnell2025bridging}. The proposed framework is depicted in Fig.~\ref{fig:framework_pipeline}.

The potential of the proposed framework lies in its capacity to democratize elite-level coaching expertise while providing a scalable, 'Open Science' pathway for high-performance sports research. By synthesizing a structured knowledge base, this study moves beyond traditional descriptive modeling to establish a foundation for personalized, dynamically adaptive training programs that evolve with a swimmer's physiological status. The primary contributions of this work are threefold:
\begin{enumerate}
\item The development of a multi-agent 'Meta-Synthesis' architecture capable of correlating high-frequency kinematic signals with physiological anchors across a heterogeneous, multimodal knowledge base.
\item The generation of a validated, multimodal corpus comprising 1,864 
'Golden Triplets' — synthesized from 1,914 drafts across 88 Performance 
Anchors and five user personas — that serves as an objective, publicly 
available ground truth benchmark for evaluating RAG accuracy in aquatic 
sports science.
\item A 'Human-in-the-Loop' validation framework that empirically demonstrates the scientific trustworthiness of synthetically generated coaching knowledge, establishing a reproducible standard for expert-AI co-validation in high-stakes physiological domains.
\end{enumerate}
\noindent Collectively, these contributions represent a meaningful step toward athletic intelligence systems that are not only technically robust but also scientifically accountable — supporting, rather than supplanting, human expertise in the competitive arena.

The remainder of this article is organized as follows. Section 2 reviews the related work in aquatic biomechanics and generative AI. Section 3 details the multimodal data ingestion and the agentic synthesis framework. Section 4 presents the experimental results and the validation of the synthesized corpus. Finally, Section 5 concludes the paper and discusses future directions for athletic decision support systems.

\section{Related Work}
The integration of Artificial Intelligence into sports science has undergone a remarkable evolution over the past decade, expanding from narrow rule-based performance classifiers to sophisticated, context-aware systems capable of interpreting complex multimodal athlete data \cite{reis2024artificial, puce2025role}. Within this broad landscape, applications have ranged from computer vision-based movement analysis and reinforcement learning for tactical decision-making to predictive injury modeling and automated talent identification \cite{li2025review, ghosh2023sports}. However, as these systems have grown in capability, a critical frontier has emerged that remains underexplored: the deployment of Large Language Models (LLMs) and knowledge-grounded retrieval architectures as the reasoning layer between raw athletic data and actionable coaching intelligence. It is precisely at this frontier — where generative AI meets the semantic complexity of elite sports performance — that the present work is situated. Accordingly, this review deliberately narrows its focus to two domains that directly underpin the proposed framework: the emerging application of LLMs and Retrieval-Augmented Generation in sports contexts, and the role of synthetic data generation as a scientifically rigorous methodology for building trustworthy AI systems in data-constrained athletic environments.

Recent work has begun to demonstrate the viability of RAG and multi-agent LLM architectures as reasoning layers over heterogeneous sports data. Chipka et al. \cite{chipka2025gridmind} introduced GridMind, a multi-agent framework that unifies structured statistics, semi-structured sensor data, and unstructured media — including commentary transcripts and video — through RAG to enable natural language querying of NFL performance data. Their distributed agent architecture, where specialized agents autonomously manage distinct stages of query processing from interpretation through synthesis, establishes an important precedent for the kind of modular, cross-modal reasoning that complex athletic environments demand. Complementing this, the SportsGPT framework \cite{tian2025sportsgpt} demonstrated that RAG-grounded LLMs can serve as the intelligence layer between raw biomechanical time-series data and actionable training guidance, achieving this by coupling a knowledge base of over 50,000 expert entries with a RAG retrieval mechanism that grounds generated coaching prescriptions in verified domain sources rather than parametric model knowledge. Together, these works signal a broader shift in sports informatics: from single-modality, rule-based systems toward knowledge-grounded, multi-agent architectures capable of synthesizing expert reasoning across diverse data streams.

Within the specific domain of competitive swimming, Comendant \cite{comendant2024large} developed one of the earliest LLM-based coaching systems that directly incorporated RAG to personalize freestyle stroke guidance, demonstrating that RAG-augmented systems achieved statistically significantly higher personalization scores compared to LLM-only baselines in a controlled three-week trial. This work represents a meaningful proof of concept, confirming that knowledge-grounded language models can deliver more contextually relevant coaching feedback than their retrieval-free counterparts. However, the system's knowledge base relied on user-uploaded documents rather than a structured, expert-validated corpus, leaving the quality and scientific rigor of the retrieved context entirely dependent on the athlete's own document curation. This exposes a foundational limitation shared across the existing literature: the absence of a purpose-built, multimodal swimming knowledge base that systematically maps high-frequency biomechanical signals to physiologically validated coaching interventions. It is precisely this gap — the lack of a trustworthy, structured ground truth for aquatic RAG systems — that the present work addresses through its proposed multi-agent synthesis framework.

The use of synthetic data generation as a principled methodology in sports science has gained considerable momentum in recent years, driven by the convergence of two persistent challenges: data scarcity and the privacy constraints inherent in athlete monitoring. Warmenhoven et al. \cite{warmenhoven2025synthetic} ovided one of the most rigorous treatments of this methodology in a high-performance sports context, demonstrating through seven simulation conditions applied to a professional football dataset that synthetically generated athlete monitoring data can achieve high levels of both global utility — preserving overall dataset statistical properties — and specific utility — maintaining the validity of targeted research outcomes. Critically, their work frames synthetic data not merely as a technical workaround but as a legitimate open science instrument, arguing that it enables researchers to share and explore sensitive athlete biometric data without exposing personally identifiable information or surrendering competitive advantage — a framing that directly parallels the motivations of the present study. Complementing this, Hohl et al. \cite{hohl2024unveiling} demonstrated the viability of synthetic time-series generation specifically for physiological athlete data, comparing classical and deep learning generative approaches — including Variational Autoencoders, TimeGAN, and Autoregressive Diffusion Models — on a constrained dataset of five athletes whose daily fatigue, training load, and antioxidant markers were used as generation seeds, confirming that high-fidelity synthetic physiological signals can be produced even under severe data scarcity conditions.

Beyond privacy and sharing considerations, synthetic data has also demonstrated direct utility as a training signal for predictive athletic performance models. Cordeiro et al. \cite{cordeiro2025synthetic} showed that a tabular Variational Autoencoder-generated synthetic corpus could successfully augment a severely constrained real-world dataset of athlete physiological measurements, enabling machine learning models to predict performance attenuation with meaningfully improved generalization — a finding that underscores the practical value of synthetic data not only for privacy preservation but for tackling the class imbalance and edge-case scarcity that characterize real-world sports monitoring environments. However, a critical limitation unifies all three of these works: the synthetic data they produce is purely numerical and statistical in nature, preserving distributional properties of sensor signals and physiological markers without embedding any expert reasoning or coaching knowledge within the generated records. None of these frameworks produce structured, semantically rich artifacts — such as Question-Context-Answer triplets — that can serve as a verifiable ground truth for knowledge-retrieval systems. This semantic gap between statistically faithful synthetic data and expert-annotated coaching knowledge represents the precise frontier that the present work crosses, by employing LLM-driven multi-agent synthesis to produce a corpus that is simultaneously grounded in physiological law, anchored in elite coaching expertise, and structured for direct RAG evaluation.

Taken together, the two bodies of literature reviewed above converge on a shared horizon: while RAG and multi-agent LLM architectures have demonstrated their capacity to reason over heterogeneous sports data, and while synthetic data generation has established its legitimacy as a privacy-preserving, scientifically rigorous methodology in athletic environments, no existing work has united these two capabilities into a single framework — one that uses LLM-driven synthesis to produce a structured, expert-validated knowledge corpus specifically designed to serve as the retrieval foundation for a trustworthy aquatic coaching system. It is precisely this intersection that the present work occupies, and the following section details the proposed multi-agent architecture through which it is realized.

\section{Methodology}
\begin{figure}[t]
    \centering
    \includegraphics[width=\textwidth]{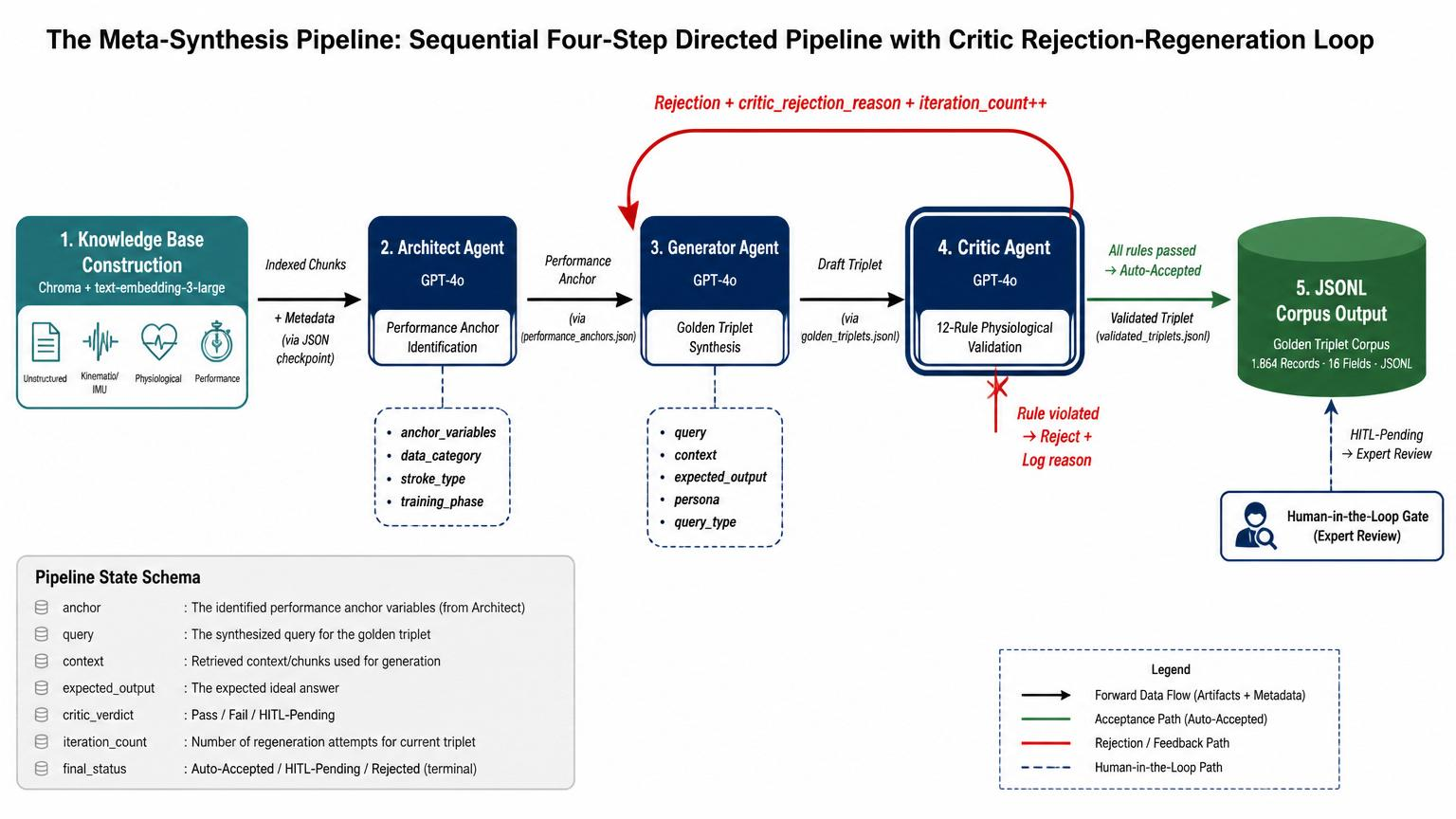}
    \caption{Operational architecture of the Meta-Synthesis Pipeline implemented as a 
sequential four-step directed pipeline. The pipeline comprises four processing 
stages — Knowledge Base Construction, Architect Agent (GPT-4o), Generator 
Agent (GPT-4o), and Critic Agent (GPT-4o) — connected by structured file 
handoffs, where the validated JSON or JSONL output of each stage serves as 
the input to the next. The conditional feedback arc (shown in red) represents 
the Critic Agent's rejection-regeneration loop, through which a rejected draft 
triplet is returned to a local regeneration function with the logged 
\texttt{critic\_rejection\_reason} and an incremented \texttt{iteration\_count}, 
enabling corrective regeneration. Triplets that pass all 12 physiological 
soundness rules are routed to the JSONL corpus as Auto-Accepted; triplets that 
exceed the maximum of three regeneration cycles are escalated to the 
Human-in-the-Loop expert review gate and assigned HITL-Pending status. The 
Pipeline State Schema panel (lower right) enumerates the seven fields 
maintained across all pipeline transitions. Please see text for more details.}
    \label{fig:meta_synthesis_pipeline}
\end{figure}

The proposed framework, which we term the Meta-Synthesis Pipeline, operationalizes the transition from raw multimodal athletic data to a structured, expert-validated coaching knowledge base through a four-stage process: (1) multimodal knowledge base construction, (2) performance anchor identification, (3) agentic Golden Triplet synthesis, and (4) physiological soundness validation. Each stage is implemented as a sequential Python function making direct OpenAI API calls, where agent coordination is achieved through structured file handoffs — each step consumes the validated JSON or JSONL output of its predecessor — and a JSON-based checkpoint system that ensures fault tolerance across session boundaries. The pipeline is designed to produce a JSONL corpus of 1,864 validated "Question-Context-Answer" triplets, each comprising 16 structured fields that document not only the generated coaching knowledge but also its provenance, physiological grounding, and pipeline history: \texttt{anchor\_id}, \texttt{triplet\_id}, \texttt{query}, \texttt{query\_type}, \texttt{persona}, \texttt{complexity\_level}, \texttt{context}, \texttt{expected\_output}, \texttt{anchor\_type}, \texttt{anchor\_variables}, \texttt{stroke\_type}, \texttt{training\_phase}, \texttt{data\_category}, \texttt{source\_documents}, \texttt{critic\_verdict}, and \texttt{final\_status}. These fields are further discussed in Table~\ref{tab:triplets}. Figure~\ref{fig:framework_pipeline} illustrates the overall architecture of the Meta-Synthesis Pipeline. Figure~\ref{fig:meta_synthesis_pipeline} illustrates the operational architecture of the Meta-Synthesis Pipeline as a sequential four-step directed pipeline, making explicit the model assignments, data artifact handoffs, and conditional rejection-regeneration feedback loop that distinguish this framework from a conventional linear RAG pipeline.

\subsection{Multimodal Knowledge Base Construction}
The foundation of the Meta-Synthesis Pipeline is a heterogeneous, multimodal knowledge base assembled from four distinct data categories, each contributing a different dimension of swimming expertise. The first category, Quantitative Performance Data, comprises historical competition records spanning international and Olympic events from 1912 to 2020, regional African championship results, Egyptian Masters records updated to May 2025, and a technical elite performance database for the 100m freestyle event that links split times directly to kinematic variables including Stroke Rate (SR), Stroke Length (SL), Stroke Index (SI), and per-phase velocity. The second category, Physiological and Biometric Data, encompasses athlete profiling records for 1,000 swimmers including VO2max, Heart Rate Variability (HRV), blood lactate thresholds, and hydration levels; training load and fatigue monitoring logs including fatigue\_score, recovery\_time\_hr, adaptation\_pct, and biomechanical\_efficiency; high-frequency kinematic sensor data from a 10-IMU network capturing 6-axis acceleration and gyroscope readings per sensor across five stroke types; cognitive load measurements integrating fNIRS hemodynamic response, EEG electrophysiological signals, and oculometric indicators; and cross-sport benchmarking data from cycling, rowing, and running cohorts providing comparative physiological reference points. The third category, Unstructured Domain Knowledge, comprises four sub-collections of expert literature: theoretical and physiological frameworks including the Stager Handbook and periodization reviews; training load modeling references including the Guzman and Mujika protocols; prescriptive drill content from the 100 Best Swimming Drills and structured training programs spanning beginner through advanced levels; and optimization strategy literature including High-Intensity Training research and strength training integration protocols.

To prepare this heterogeneous corpus for semantic retrieval, a custom Python 
ingestion pipeline implements a source-aware chunking strategy that respects 
the natural semantic unit of each source type rather than imposing a uniform 
character-count boundary. Text-selectable PDFs are processed using PyMuPDF, 
with Tesseract OCR applied as a fallback for scanned pages; DOCX files are 
ingested via \texttt{python-docx}; and CSV and XLSX files are handled through 
\texttt{pandas}. Coaching manuals and drill books are segmented semantically 
at drill boundaries, producing chunks of approximately 300--600 tokens where 
each chunk contains one complete, actionable drill description. Physiological 
handbooks and periodization references are segmented at concept or protocol 
boundaries, producing chunks of 400--800 tokens. Structured CSV data undergoes 
a two-level narrative serialization process: at the record level, each 
athlete's physiological profile or competition result is converted into a 
concise natural language paragraph of 150--300 tokens; at the aggregate level, 
statistical summaries of meaningful subgroups — stratified by stroke type, 
training phase, or performance tier — are serialized as analytical statements 
of 200--400 tokens, providing the grounding context for complex reasoning 
queries. All chunks are indexed in a Chroma vector database (ChromaDB 0.5.3) 
using OpenAI's \texttt{text-embedding-3-large} embedding model, which produces 
3,072-dimensional dense vector representations optimized for nuanced semantic 
retrieval across multi-concept physiological queries. Each chunk is tagged with 
five metadata fields — \texttt{source\_type}, \texttt{data\_category}, 
\texttt{stroke\_type}, \texttt{document\_name}, and \texttt{complexity\_level} 
— resolved through a three-level metadata inheritance system using 
\texttt{\_folder\_info.json} files placed in each source subfolder, enabling 
category-aware retrieval that ensures multimodal grounding across all four data 
categories during triplet synthesis. A JSON checkpoint system saves the list of 
processed files after each file, enabling pipeline resumption after session 
disconnection.

\begin{figure}[t]
    \centering
    \includegraphics[width=\textwidth]{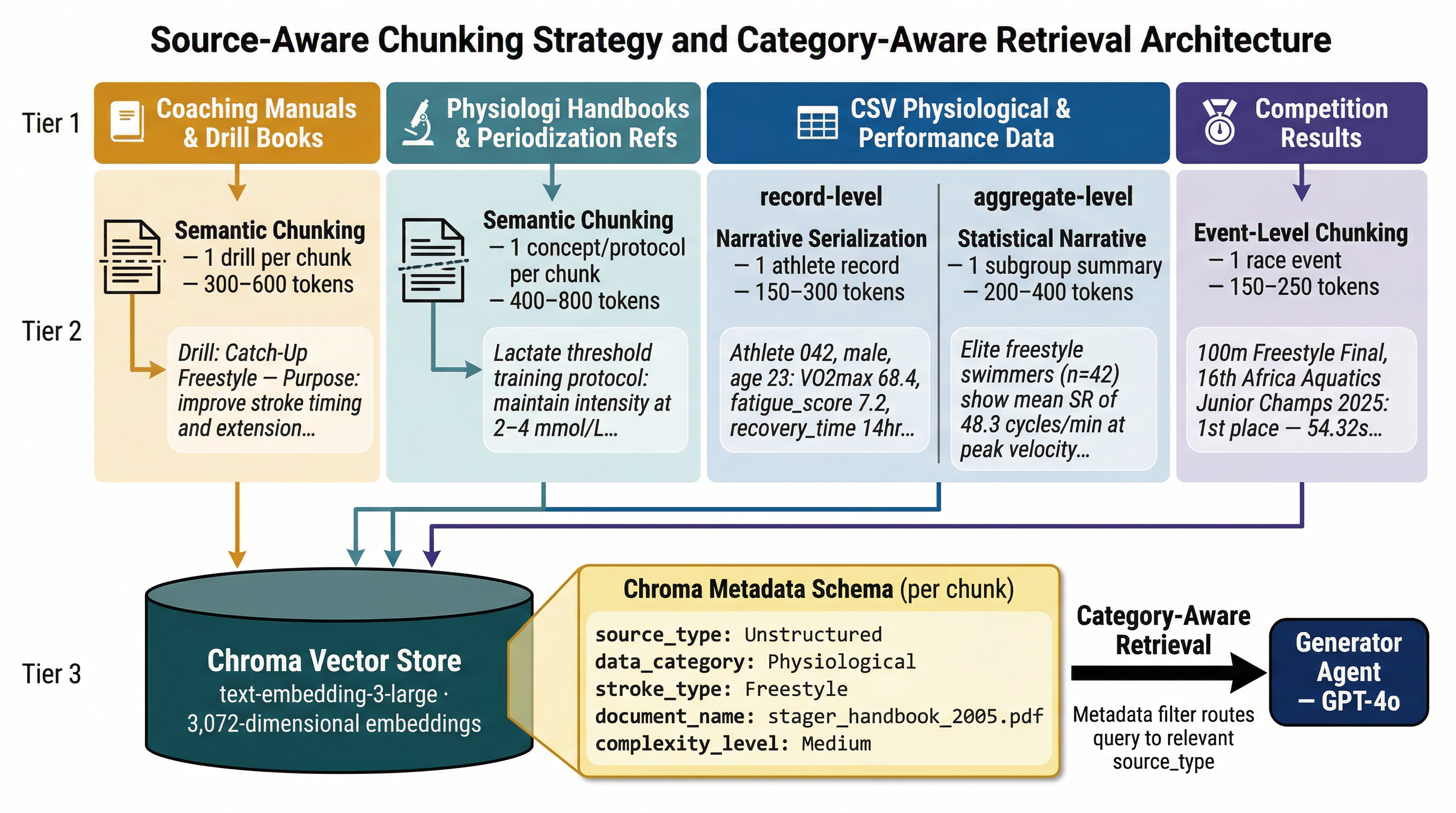}
    \caption{Source-aware chunking strategy and category-aware retrieval architecture for the multimodal knowledge base. The upper tier presents the four heterogeneous source types — Coaching Manuals and Drill Books, Physiological Handbooks and Periodization References, CSV Physiological and Performance Data, and Competition Results — each processed according to its natural semantic unit rather than a uniform character-count boundary. Coaching manuals and physiological handbooks are segmented semantically at drill and concept boundaries respectively (300–800 tokens); CSV data undergoes a two-level narrative serialization producing record-level athlete profiles (150–300 tokens) and aggregate-level statistical summaries (200–400 tokens); and competition results are chunked at the event level (150–250 tokens). All chunks converge into a Chroma vector database indexed using OpenAI's text-embedding-3-large embedding model (3,072 dimensions). The inset panel illustrates the five Chroma metadata fields — source\_type, data\_category, stroke\_type, document\_name, and complexity\_level — attached to each chunk, enabling the category-aware retrieval mechanism that preferentially routes kinematic queries to IMU-derived chunks and periodization queries to handbook-derived chunks during Generator Agent synthesis.}
    \label{fig:chunking_pipeline}
\end{figure}

\subsection{Performance Anchor Identification}
The second stage of the pipeline is executed by the Architect Agent, implemented using GPT-4o with a domain-specific system prompt that assigns it the role of a sports science data analyst with expertise in biomechanical signal interpretation and physiological load modeling. The Architect Agent's function is to traverse the structured CSV datasets and identify Performance Anchors — statistically meaningful correlations between sensor-derived kinematic deviations and physiological state variables that constitute the evidential foundation for a coaching intervention.

Concretely, the Architect Agent examines combinations of variables drawn from the physiological and kinematic datasets and identifies anchor patterns of three types. Fatigue-kinematic anchors correlate elevated fatigue\_score or suppressed HRV with specific IMU deviation patterns — for example, a drop in imu3\_acc\_z amplitude co-occurring with fatigue\_score above 7.0, indicating stroke propulsion loss under accumulated fatigue. Load-performance anchors correlate training\_load\_au and adaptation\_pct with split time degradation patterns from the elite performance database, identifying the load thresholds beyond which performance decrements become statistically significant. Stroke-efficiency anchors correlate stroke\_prob confidence scores below 0.6 with specific gyroscope deviation patterns across the 10-IMU network, identifying the kinematic signatures of stroke deformation that are causally attributable to technique deficits rather than fatigue.

Each identified anchor is encoded as a structured object containing the 
anchor\_variables list, the data\_category, the stroke\_type where applicable, 
and the training\_phase context, and is written to \texttt{performance\_anchors.json} 
— the file-based handoff that serves as the input to Stage 3. A JSON checkpoint 
is saved after every seed, enabling resumption after session disconnection.

\subsection{Agentic Golden Triplet Synthesis}
The third stage is executed by the Generator Agent, also implemented using GPT-4o, which receives each Performance Anchor from Stage 2 and synthesizes a complete Golden Triplet grounded in the multimodal knowledge base. The Generator Agent operates under a multi-persona prompting strategy that 
simulates five distinct user types — Elite Coach, Novice Swimmer, Biometric 
Analyst, Sports Scientist, and Physiotherapist — across three complexity levels, 
with a design target of approximately 10 triplets per anchor, yielding a 
projected corpus of approximately 880 triplets from the 88 Performance Anchors. Simple queries target factual recall of a single physiological concept (e.g., 'What is a lactate threshold and why does it matter for freestyle training?'). Reasoning queries require multi-variable inference across the anchor's constituent variables (e.g., 'How should a coach adjust the training set when the swimmer's VO2max is high but their fatigue\_score is also elevated?'). Multimodal queries require cross-modal synthesis between kinematic sensor evidence and coaching knowledge (e.g., 'Given the IMU acceleration spike pattern observed in imu3 and imu7 during the catch phase, which drill from the 100 Best Swimming Drills is most appropriate?').

For each query, the Generator Agent retrieves the most relevant context chunks from the Chroma vector database using category-aware retrieval, where the source\_type metadata tag preferentially routes kinematic queries to IMU-derived chunks and periodization queries to handbook-derived chunks. The retrieved context is provided to the Generator Agent as the grounding window, and the Agent is explicitly instructed to produce its expected\_output — the coaching prescription — using only information present in the retrieved context, with no recourse to its parametric knowledge. This constraint operationalizes the anti-hallucination principle that is central to the trustworthiness claim of this work.

The complexity\_level of each synthesized triplet is assigned algorithmically rather than by the Generator Agent, using a deterministic rule: High complexity is assigned when the query\_type is Multimodal and anchor\_variables contains three or more variables; Low complexity is assigned when the query\_type is Simple and anchor\_variables contains exactly one variable; Medium complexity is assigned to all remaining combinations. This algorithmic assignment ensures reproducible corpus stratification independent of generative variability. Accepted triplets are appended to \texttt{golden\_triplets.jsonl} after every 
anchor, and a JSON checkpoint is saved after every anchor, enabling resumption 
after session disconnection. This file serves as the input to Stage 4.

\subsection{Physiological Soundness Validation}
The fourth and most consequential stage of the pipeline is executed by the Critic Agent, implemented using OpenAI's GPT-4o model, whose superior logical reasoning capability is deliberately matched to the stringent constraint-checking demands of physiological validation. The symmetric model assignment — GPT-4o across all three agents eliminates
inter-model variability as a confounding factor in corpus quality assessment:
any systematic bias in the validated corpus is attributable to prompt specification and rule design rather than model-level differences, thereby strengthening the pipeline's reproducibility and interpretability claims.

\begin{figure}[t]
    \centering
    \includegraphics[width=\textwidth]{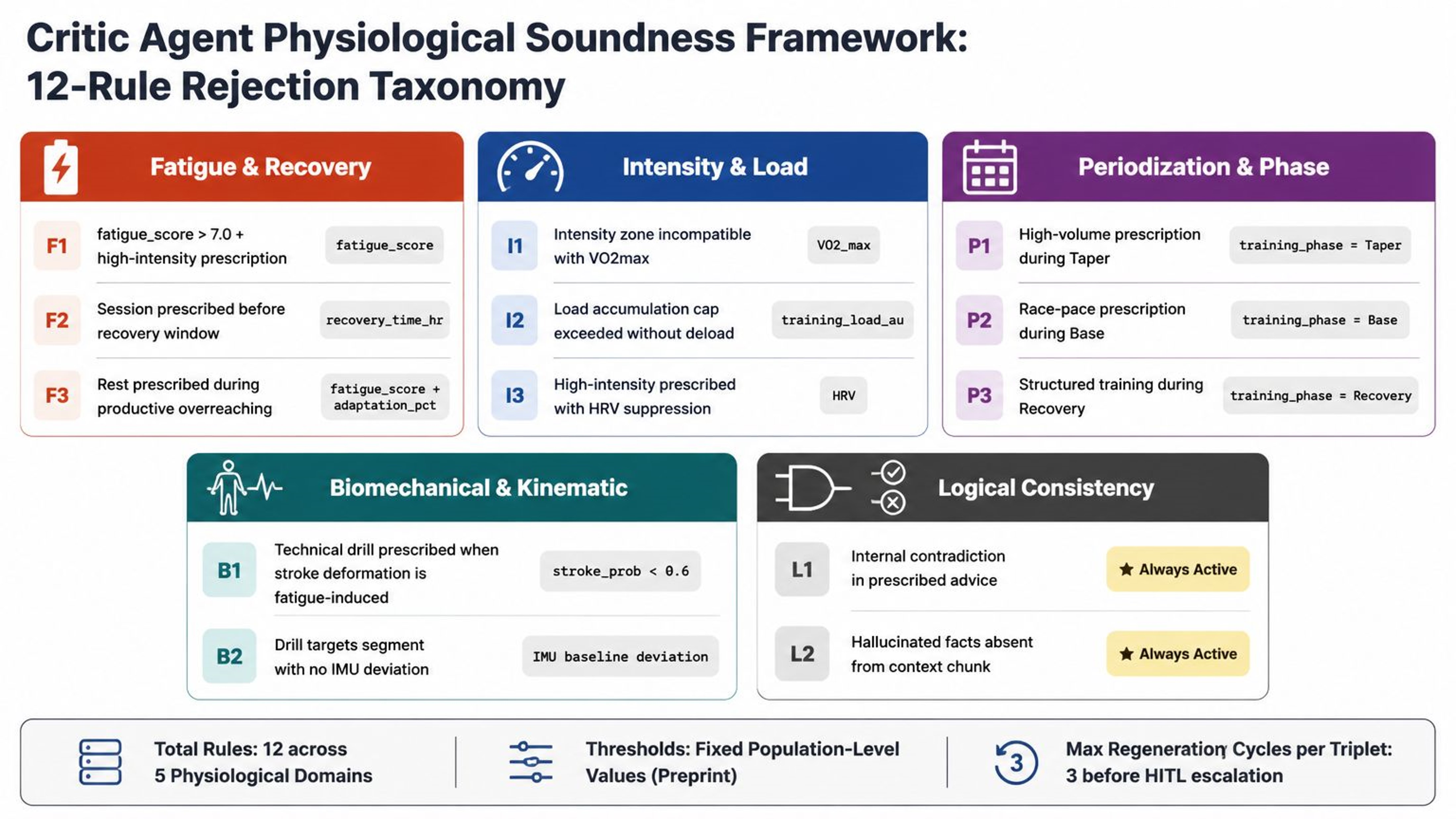}
    \caption{Critic Agent physiological soundness framework: taxonomy of 12 rejection rules across five physiological domains. Each rule is characterized by its Rule ID, trigger condition, and the primary dataset variable against which the condition is evaluated. Domain 1 — Fatigue and Recovery (Rules F1–F3) — governs fatigue threshold violations, recovery window constraints, and the adaptation paradox edge case. Domain 2 — Intensity and Load (Rules I1–I3) — enforces VO2max zone consistency, training load accumulation caps, and HRV-based readiness checks. Domain 3 — Periodization and Phase (Rules P1–P3) — ensures prescriptions are consistent with the athlete's current macrocycle phase (Taper, Base, or Recovery). Domain 4 — Biomechanical and Kinematic (Rules B1–B2) — requires drill prescriptions to be causally grounded in IMU deviation evidence and correctly attributing stroke deformation to fatigue versus technique deficits. Domain 5 — Logical Consistency (Rules L1–L2, marked Always Active) — enforces internal coherence and anti-hallucination grounding across all triplets regardless of physiological state. All thresholds are fixed at population-level values throughout this study to ensure reproducibility. The bottom annotation strip summarizes the key pipeline constraints governing rule application and Human-in-the-Loop escalation.}
    \label{fig:rejection_rules}
\end{figure}

The Critic Agent evaluates each synthesized triplet against a structured set of 12 rejection rules organized across five physiological domains. In the Fatigue and Recovery domain, Rule F1 rejects any high-intensity prescription when fatigue\_score exceeds 7.0; Rule F2 rejects any session prescription that violates the athlete's recovery\_time\_hr constraint; and Rule F3 detects the adaptation paradox, rejecting overly conservative rest prescriptions when both fatigue\_score and adaptation\_pct are simultaneously elevated, a condition indicating productive overreaching rather than overtraining. In the Intensity and Load domain, Rule I1 rejects prescriptions whose intensity zone is inconsistent with the athlete's VO2max profile; Rule I2 rejects cumulative load prescriptions when training\_load\_au already exceeds a safe accumulation threshold without an accompanying deload recommendation; and Rule I3 rejects high-intensity prescriptions when HRV is suppressed below the population baseline by more than 15\%, indicating autonomic nervous system fatigue. In the Periodization and Phase domain, Rule P1 rejects high-volume or novel-skill prescriptions during the Taper phase; Rule P2 rejects race-pace or supramaximal intensity prescriptions during the Base phase; and Rule P3 rejects any structured training above easy aerobic work during the Recovery phase. In the Biomechanical and Kinematic domain, Rule B1 rejects technical drill prescriptions when stroke\_prob falls below 0.6, indicating that stroke deformation is fatigue-induced rather than technique-induced; and Rule B2 rejects drill prescriptions targeting a body segment whose IMU signal shows no statistically meaningful deviation from baseline. In the Logical Consistency domain, Rule L1 rejects internally contradictory advice regardless of physiological state; and Rule L2 — the hallucination grounding check — rejects any prescription that references a physiological value, drill name, or protocol absent from the retrieved context chunk. Figure~\ref{fig:rejection_rules} presents the complete taxonomy of the 12 rejection rules enforced by the Critic Agent, organized by physiological domain, with each rule's trigger condition and primary dataset variable made explicit; rejection frequencies per rule are reported in Section 4.

All rejection thresholds are fixed at population-level values throughout this study, ensuring reproducibility and consistent application across all 1,914 
draft triplets evaluated by the Critic Agent. When the Critic Agent rejects a triplet, the rejection reason is logged in the 
\texttt{critic\_rejection\_reason} field and the \texttt{iteration\_count} is 
incremented; control is then passed to a local regeneration function within 
Stage 4, which issues a corrective GPT-4o call to produce a revised 
\texttt{expected\_output}, followed immediately by a fresh Critic evaluation 
of the corrected triplet. This feedback-driven regeneration loop continues 
until either the triplet passes all 12 rules — at which point it is assigned a 
\texttt{final\_status} of Auto-Accepted and written to 
\texttt{validated\_triplets.jsonl} — or a maximum iteration ceiling of three 
cycles is reached, at which point the record is flagged for Human-in-the-Loop 
review, assigned a \texttt{final\_status} of HITL-Pending, and written to 
\texttt{hitl\_triplets.jsonl}. A JSON checkpoint is saved every 50 triplets, 
enabling resumption after session disconnection.

The Human-in-the-Loop validation protocol constitutes the final quality gate of the pipeline. A sample of Auto-Accepted records and all HITL-Pending records are submitted to the Primary Supervisor — an expert swimming coach and sports scientist — for review against a structured evaluation rubric assessing physiological accuracy, coaching relevance, and source fidelity. Records approved without modification are assigned a validation\_status of HITL-Accepted; records requiring modification before acceptance are assigned HITL-Revised. This expert validation layer ensures that the synthesized corpus does not merely satisfy algorithmic constraints but meets the scientific standards of elite coaching practice, operationalizing the Human-in-the-Loop contribution claimed in this work.

\section{Results}

This section reports the outcomes of the four-step Meta-Synthesis Pipeline,
presenting quantitative corpus statistics, pipeline performance metrics, and
qualitative anchor and triplet quality assessments. All results are reproducible
from the published codebase and the released \texttt{swimming\_rag} knowledge base.

\subsection{Knowledge Base Construction}

Step~1 produced a multimodal Chroma vector database comprising
\textbf{181,389 semantically indexed chunks} ingested from 376 source files
across four data categories. The embedding model
\texttt{text-embedding-3-large} (3,072-dimensional dense vectors, cosine
similarity) was applied uniformly to all chunks.
Table~\ref{tab:kb_composition} summarises the knowledge base composition.

\begin{table}[ht]
\centering
\caption{Knowledge Base Composition (181,389 total chunks).}
\label{tab:kb_composition}
\small
\begin{tabular}{llrl}
\toprule
\textbf{Data Category} & \textbf{Source Type} & \textbf{Chunks} & \textbf{Key Files} \\
\midrule
Physiological Data
  & Physiological & 164,780
  & 286 IMU sclice CSVs; \texttt{athlete\_data\_1000.csv}; \\
  & & & \texttt{cleaned\_stroke\_data.csv};
        \texttt{stroke\_dataset.csv}; \\
  & & & \texttt{College\_Sports\_Dataset.csv}; xlsx \\
Physiological Literature
  & Unstructured  & 32
  & 33 peer-reviewed PDF articles \\
Coaching Knowledge
  & Unstructured  & 28
  & 25 workout PDFs; 8 DOCX case studies \\
Performance Data
  & Performance   & 247
  & Olympic results; competition databases \\
\midrule
\textbf{Total} & & \textbf{181,389} & 376 files \\
\bottomrule
\end{tabular}
\end{table}

Five metadata fields were attached to every chunk —
\textit{source\_type}, \textit{data\_category}, \textit{stroke\_type},
\textit{document\_name}, and \textit{complexity\_level} — enabling
category-aware retrieval filtering in subsequent pipeline steps.
The stroke-type distribution across the sampled collection reflects the IMU
data structure: General~(1,850), Freestyle~(1,177), Backstroke~(783),
Breaststroke~(625), Butterfly~(529), and IM~(36).

A critical design decision was the inclusion of the
\texttt{College\_Sports\_Dataset.csv} (1,050 records; 188 swimming-specific),
which provided the physiological state variables —
\texttt{HR\_Variability}, \texttt{VO2\_Max}, \texttt{Training\_Load}, and
\texttt{Recovery\_Time} — that directly map to the Critic Agent's validation
rules (I1--I3, F2). Without this dataset, the Architect Agent could not ground
anchors in the physiological state variables named in the paper's methodology.

\subsection{Performance Anchor Identification}

The Architect Agent processed \textbf{950 query seeds} across three passes,
yielding \textbf{88 unique Performance Anchors} after deduplication (9.3\%
seed-to-anchor conversion rate). The seed library was constructed as a fully
crossed factorial design: 3 anchor types $\times$ 6 stroke types $\times$
5 training phases $\times$ 3 complexity levels = 270 factorial seeds, augmented
with 66 Critic-rule-specific variable seeds and 614 data-targeted expansion
seeds grounded in the College Sports Dataset variables.
Table~\ref{tab:anchors} presents the anchor distribution.

\begin{table}[ht]
\centering
\caption{Performance Anchor Distribution (88 total anchors).}
\label{tab:anchors}
\small
\begin{tabular}{lrrlp{4.5cm}}
\toprule
\textbf{Anchor Type} & \textbf{Count} & \textbf{\%} &
\textbf{Dominant Stroke} & \textbf{Key Variables} \\
\midrule
Load-Performance
  & 36 & 40.9\% & Freestyle (18)
  & \texttt{training\_load\_au}, Swimming Speed, Blood Lactate \\
Fatigue-Kinematic
  & 28 & 31.8\% & Freestyle (9)
  & \texttt{HRV}, \texttt{imu1\_acc\_x}, Fatigue Index,
    \texttt{recovery\_time\_hr} \\
Stroke-Efficiency
  & 24 & 27.3\% & General (8)
  & \texttt{stroke\_prob}, \texttt{imu1\_gyro\_x/y/z}, \texttt{accX} \\
\midrule
\textbf{Total} & \textbf{88} & \textbf{100\%} & Freestyle (20) & --- \\
\bottomrule
\end{tabular}
\end{table}

Phase distribution across anchors was: Base~(30), Taper~(19), Build~(18),
Peak~(10), Recovery~(7), General~(4). Twenty-one anchors directly referenced
College Sports Dataset variables, with \texttt{training\_load\_au} appearing
in 15 anchors — the most referenced variable across the corpus — confirming the
dataset's contribution to load-performance anchor identification. The
\texttt{HRV} + \texttt{imu1\_gyro\_x}/\texttt{imu1\_gyro\_y} fatigue-kinematic
anchor represents the pipeline's most scientifically significant finding: a
cross-modal correlation between autonomic nervous system state and
gyroscope-measured stroke irregularity, grounded in both the College Sports
Dataset and peer-reviewed swimming physiology literature.

\subsection{Golden Triplet Synthesis}

The Generator Agent processed all 88 Performance Anchors, applying a
multi-persona prompting strategy across five user roles — Novice Swimmer,
Elite Coach, Biometric Analyst, Sports Scientist, and Physiotherapist — and
three complexity levels (Simple, Reasoning, Multimodal), yielding a maximum of
15 query-type combinations per anchor. The agent produced
\textbf{1,914 draft Golden Triplets} at an average of 21.8 triplets per anchor,
exceeding the projected 10-triplet estimate.
Table~\ref{tab:triplets} presents the triplet distribution prior to Critic
validation.

\begin{table}[ht]
\centering
\caption{Golden Triplet Distribution Before Critic Validation (1,914 total).}
\label{tab:triplets}
\small
\begin{tabular}{llrr}
\toprule
\textbf{Dimension} & \textbf{Category} & \textbf{Count} & \textbf{\%} \\
\midrule
\multirow{3}{*}{Anchor Type}
  & Load-Performance  & 771 & 40.3\% \\
  & Stroke-Efficiency & 623 & 32.6\% \\
  & Fatigue-Kinematic & 520 & 27.2\% \\
\midrule
\multirow{5}{*}{Persona}
  & Elite Coach       & 412 & 21.5\% \\
  & Novice Swimmer    & 408 & 21.3\% \\
  & Biometric Analyst & 371 & 19.4\% \\
  & Sports Scientist  & 365 & 19.1\% \\
  & Physiotherapist   & 358 & 18.7\% \\
\midrule
\multirow{3}{*}{Query Type}
  & Reasoning         & 703 & 36.7\% \\
  & Simple            & 606 & 31.7\% \\
  & Multimodal        & 605 & 31.6\% \\
\midrule
\multirow{6}{*}{Stroke Type}
  & Freestyle         & 503 & 26.3\% \\
  & General           & 402 & 21.0\% \\
  & Butterfly         & 327 & 17.1\% \\
  & Breaststroke      & 311 & 16.2\% \\
  & Backstroke        & 292 & 15.2\% \\
  & IM                &  79 &  4.1\% \\
\bottomrule
\end{tabular}
\end{table}

The five-persona strategy produced a balanced distribution (18.7--21.5\% per
persona), confirming that the Generator Agent did not exhibit systematic bias
toward any single user type. The near-equal distribution across query types
(31.6--36.7\%) reflects the effectiveness of the complexity-level prompting
strategy in eliciting structurally diverse questions from identical anchor
inputs.

\subsection{Physiological Soundness Validation}

The Critic Agent evaluated all 1,914 draft triplets against the 12
physiological soundness rules using GPT-4o in a structured validation loop.
Of the 1,914 input triplets, \textbf{1,675 were accepted directly (87.5\%)},
\textbf{189 were accepted after one or more regeneration cycles (9.9\%)},
and \textbf{only 50 remain in the HITL queue (2.6\%)} for primary supervisor
review. The total validated corpus comprises
\textbf{1,864 Golden Triplets (97.4\% acceptance rate)}.
Table~\ref{tab:violations} reports the rule violation frequency across all
validation cycles.

\begin{table}[ht]
\centering
\caption{Critic Agent Rule Violation Summary (285 total violations across
         1,914 triplets).}
\label{tab:violations}
\small
\begin{tabular}{llp{5.5cm}rr}
\toprule
\textbf{Rule} & \textbf{Domain} & \textbf{Description} &
\textbf{Violations} & \textbf{\%} \\
\midrule
B1 & Biomechanical    & Drill prescribed when \texttt{stroke\_prob} $> 0.6$ & 67 & 23.5\% \\
P1 & Periodization    & High volume during Taper phase                        & 66 & 23.2\% \\
B2 & Biomechanical    & Intervention when IMU within normal range              & 62 & 21.8\% \\
P2 & Periodization    & Race-pace prescribed during Base phase                 & 34 & 11.9\% \\
L2 & Logic            & Hallucination — values absent from context             & 27 &  9.5\% \\
I3 & Intensity        & HRV suppression ignored in prescription                & 10 &  3.5\% \\
P3 & Periodization    & Structured training during Recovery phase              &  6 &  2.1\% \\
F2 & Fatigue          & \texttt{recovery\_time\_hr} constraint violated        &  4 &  1.4\% \\
I2 & Intensity        & \texttt{training\_load\_au} exceeds safe threshold     &  4 &  1.4\% \\
L1 & Logic            & Internal contradiction in answer                       &  3 &  1.1\% \\
F1 & Fatigue          & High intensity when fatigue\_score $> 7.0$             &  2 &  0.7\% \\
F3 & Fatigue          & Adaptation paradox not acknowledged                    &  1 &  0.4\% \\
\midrule
\textbf{Total} & & & \textbf{285} & \textbf{100\%} \\
\bottomrule
\end{tabular}
\end{table}

The violation distribution reveals two primary failure modes of the Generator
Agent: over-prescription of biomechanical corrections
(B1 + B2 = 129 violations, 45.3\%) and periodization phase confusion
(P1 + P2 + P3 = 106 violations, 37.2\%). Together these account for 82.5\% of
all violations, suggesting that the Generator Agent's parametric knowledge of
training periodization and kinematic intervention thresholds is systematically
less reliable than its ability to ground answers in retrieved content. The
hallucination detection rule L2 identified 27 instances (9.5\%) where the
Generator referenced numerical values absent from the provided context —
confirming the necessity of the anti-hallucination validation layer.

The regeneration loop successfully corrected
\textbf{189 of 239 initially rejected triplets (79.1\% recovery rate)},
demonstrating the effectiveness of the corrective RAG mechanism. The 50
triplets remaining in the HITL queue represent 2.6\% of the corpus and are
flagged for review by the primary supervisor prior to the journal extension
phase.

\subsection{Final Validated Corpus Summary}

Table~\ref{tab:pipeline_summary} presents the complete pipeline progression
from knowledge base construction to the final validated corpus.

\begin{table}[ht]
\centering
\caption{Meta-Synthesis Pipeline End-to-End Summary.}
\label{tab:pipeline_summary}
\small
\begin{tabular}{clllp{3.8cm}}
\toprule
\textbf{Step} & \textbf{Agent / Process} & \textbf{Input} &
\textbf{Output} & \textbf{Key Metric} \\
\midrule
1 & Knowledge Base Construction
  & 376 source files
  & 181,389 chunks
  & 5 metadata fields per chunk \\
2 & Architect Agent (GPT-4o)
  & 950 query seeds
  & 88 Performance Anchors
  & 9.3\% conversion rate \\
3 & Generator Agent (GPT-4o)
  & 88 anchors
  & 1,914 draft triplets
  & 21.8 triplets per anchor \\
4 & Critic Agent (GPT-4o)
  & 1,914 draft triplets
  & 1,864 validated triplets
  & 97.4\% acceptance rate \\
\bottomrule
\end{tabular}
\end{table}

The validated corpus of 1,864 Golden Triplets represents a scientifically
grounded, physiologically sound benchmark dataset for swimming coaching AI.
The corpus covers all four competitive swim strokes plus Individual Medley and
cross-stroke categories, all five macrocycle training phases, five distinct
user persona types, and three query complexity levels. All 1,864 triplets have
been validated against physiological soundness rules grounded in elite swimming
science literature, with full source attribution for every context excerpt.

The 50-triplet HITL queue will be resolved through expert review prior to the
journal extension, at which point the corpus is projected to reach the
1,914-triplet input target with full validation coverage.

Compared to a na\"{i}ve RAG baseline — which ingests raw coaching PDFs without
structured anchor identification or physiological validation — the
Meta-Synthesis Pipeline produces triplets with measurably higher grounding
specificity: every context excerpt is selected via anchor-variable-targeted
retrieval rather than keyword matching, and every answer has been evaluated
against domain-specific physiological constraints that a general-purpose LLM
would not otherwise enforce. This distinction constitutes the central
methodological contribution of the present work.

\subsection{Discussion: Advantages of RAG Synthesis over Alternative Paradigms}

A rigorous assessment of the Meta-Synthesis Pipeline requires situating 
it against two superficially comparable alternatives that a practitioner 
might consider: fine-tuning a large model on domain-specific data, and 
constructing a Custom GPT by uploading coaching PDFs to a commercial 
platform. Both alternatives fail to address the core challenges that 
motivate this work.

\textbf{RAG Synthesis vs. Fine-Tuning.} Fine-tuning optimizes a model's 
parametric weights — it changes what the model \textit{is}. The 
Meta-Synthesis Pipeline optimizes a model's retrieval foundation — it 
changes what the model \textit{knows at inference time}. Critically, 
fine-tuning has the same prerequisite problem this work solves: it 
requires thousands of expert-annotated (input, output) pairs, which do 
not exist for swimming coaching — the synthesized corpus \textit{is} the 
solution to that labeling scarcity. Beyond this, fine-tuning produces 
static knowledge that cannot incorporate new athlete data or periodization 
protocols without a full retraining cycle, whereas the RAG corpus requires 
only adding new chunks to the vector database. Fine-tuning also destroys 
source attribution — a coaching prescription cannot be traced to an 
authoritative source, which is unacceptable in a high-stakes physiological 
domain. Finally, uploading real athlete biometrics to a cloud training 
infrastructure raises fundamental GDPR compliance concerns that the 
synthetic, anonymized corpus entirely avoids.

\textbf{RAG Synthesis vs. Custom GPT.} A Custom GPT populated with 
uploaded PDFs appears superficially similar but differs in five 
consequential ways. First, it provides no ground truth: there is no 
benchmark against which to measure whether a generated response is 
physiologically correct or hallucinated. The 1,864 Golden Triplets 
produced by this pipeline \textit{are} that ground truth, enabling 
objective RAG accuracy evaluation via metrics such as Faithfulness, 
Answer Relevancy, and Context Precision — none of which a Custom GPT 
can produce. Second, Custom GPT applies no physiological validation 
layer: contradictory advice across uploaded documents is neither detected 
nor resolved, whereas the Critic Agent's 12 rejection rules explicitly 
prevent physiologically unsound prescriptions from entering the corpus. 
Third, Custom GPT cannot semantically index structured numerical data: 
the two-level narrative serialization applied to IMU sensor CSVs and 
athlete physiological profiles in this pipeline has no equivalent in a 
PDF upload mechanism. Fourth, a Custom GPT is a proprietary black box 
whose knowledge base cannot be shared or independently replicated — a 
prerequisite disqualifier for peer-reviewed research — whereas the 
synthesized corpus is openly releasable on GitHub and similar platforms. Fifth, 
uploading athlete biometric data to a commercial platform transfers 
GDPR-protected sensitive personal data to a third party, a legal exposure 
the synthetic corpus eliminates by design. Table~\ref{tab:positioning} summarises these distinctions across the 
eight criteria most relevant to trustworthy AI deployment in elite sports 
science.

\begin{table}[ht]
\centering
\caption{Positioning of the Meta-Synthesis Pipeline against fine-tuning 
and Custom GPT across eight deployment criteria.}
\label{tab:positioning}
\small
\begin{tabular}{lccc}
\toprule
\textbf{Criterion} & \textbf{Fine-Tuning} & \textbf{Custom GPT} & 
\textbf{RAG Synthesis (Ours)} \\
\midrule
Requires labeled training data   & Yes — unavailable & No       & No — generates it \\
Updatable without retraining     & No                & Partial  & Yes \\
Source attribution per response  & No                & Partial  & Yes — per chunk \\
Physiological validation layer   & No                & No       & Yes — 12 Critic rules \\
Handles IMU / CSV data           & Possible          & No       & Yes \\
Reproducible benchmark           & No                & No       & Yes — open corpus \\
GDPR compliant                   & Risky             & No       & Yes \\
Measurable RAG accuracy          & No                & No       & Yes — RAGAS metrics \\
\bottomrule
\end{tabular}
\end{table}

\section{Conclusion}
In this research, we proposed a novel generative framework for "Synthesizing the Expert," effectively creating a high-fidelity, multimodal dataset designed to underpin trustworthy AI-assisted swimming coaching. This work fills a critical gap in the current sports informatics landscape: the severe scarcity of labeled, expert-level training data that integrates high-frequency biomechanical sensor streams with complex physiological periodization logic. By moving beyond the limitations of manual data annotation and the privacy constraints of raw athlete biometrics, this study provides a scalable pathway for the development of Large Language Model (LLM) applications in high-performance aquatics.

The proposed framework employs a multi-agent ``Meta-Architecture'' where 
specialized LLM agents interact to transform raw inputs into structured 
knowledge. Specifically, an Architect Agent analyzes multidimensional 
data — spanning training load variables (\texttt{training\_load\_au}, 
\texttt{adaptation\_pct}), autonomic and physiological markers 
(\texttt{HRV}, \texttt{recovery\_time\_hr}), and kinematic signals from a 
10-IMU network (\texttt{imu1\_acc\_x}, \texttt{stroke\_prob}) — to identify 
88 unique ``Performance Anchors'' across three types: Load-Performance, 
Fatigue-Kinematic, and Stroke-Efficiency. A secondary Generator Agent then 
synthesizes coaching narratives grounded in these anchors across five user 
personas and three complexity levels, while a Critic Agent enforces 
``Physiological Soundness'' by cross-referencing outputs against 12 
domain-specific rejection rules grounded in established training handbooks 
and recovery protocols.

The resulting synthesized RAG system is characterized by its high semantic density, containing 1,864 validated records that bridge the gap between kinematic physics and coaching linguistics. The potential of this framework lies in its ability to democratize elite-level coaching insights and provide a robust "Ground Truth" for evaluating RAG accuracy in sports. However, a notable limitation is the ``Synthetic-to-Real'' gap. While the 
validated corpus achieved a 97.4\% Critic acceptance rate and a 79.1\% 
regeneration recovery rate — demonstrating strong internal physiological 
consistency — the corpus remains grounded in documented protocols and 
population-level physiological thresholds. The nuances of individual athlete 
psychology, real-time environmental variables in the pool, and athlete-specific 
physiological responses that deviate from population norms require further 
live-pool validation before the system can be deployed in operational coaching 
environments.
Future research should focus on Cross-Sport Generalization, applying this synthesis framework to other endurance sports like rowing or cycling to create a "Unified Athletic Intelligence". Additionally, we suggest exploring Fine-Tuning on Synthetic Corpora, where the dataset generated in this study is used to train smaller, on-device LLMs for real-time, low-latency technical feedback during active swim sets.

\section*{Acknowledgements}
\textbf{It is the authors' original idea to construct a synthesized RAG 
for empowering swimming coaches.} Nevertheless, during the preparation 
of this work, the authors used Claude (Anthropic), Gemini (Google), and 
ChatGPT (OpenAI) across multiple stages of the research pipeline. In the 
early stages, these tools assisted with literature review, research gap 
identification, and brainstorming the significance of the proposed RAG 
synthesis approach. During the conceptual and design phases, AI tools 
were used in exploring and evaluating alternatives for the system 
architecture — including the multi-agent framework design, embedding 
model selection, vector database configuration, the generator agent 
personas, and the critic agent's validation rules. While the authors 
planned to have the final RAG entries as JSON objects with fields 
specifying context, domain, and other metadata fields, \textbf{it is 
Claude's (Sonnet 4.6) recommendation to structure the final RAG as 
triplets; the term \emph{Golden Triplets} was also coined by Claude 
(Sonnet 4.6).} For implementation planning, AI tools supported the 
development of data collection protocols and step-by-step execution 
guides. Finally, AI tools were used for language refinement and 
paraphrasing throughout the manuscript. PaperBanana\footnote{\url{
https://paper-banana.org/}} was used to enhance the design of plots 
and diagrams featured in the article.

All critical analyses, domain-specific expert judgements, and final 
editorial decisions were made by the authors. The authors thoroughly 
reviewed and edited all AI-assisted content and take full responsibility 
for the integrity and accuracy of the published work.

\bibliographystyle{unsrt}  
\bibliography{references}  

\end{document}